\documentclass[preprint,showpacs,preprintnumbers,amsmath,amssymb]{revtex4}

\usepackage{latexsym,amssymb,makeidx}
\usepackage{epsfig, color, ulem}
\usepackage{graphicx}
\usepackage{amsmath}

\begin{document}

\def\bx{{\bf x}}
\def\rmd{{\rm d}}
\def\bxS{{\bf x}_S}
\def\bxR{{\bf x}_R}
\def\setdD{\mathbb{S}}
\def\setdDR{\setdD_0}
\def\setdDA{\setdD_A}
\def\setdDB{\setdD_B}
\def\RA{R_A}
\def\RAA{R_{A}^0}
\def\RAAb{\bar R_{A}^0}
\def\RAB{R_{AB}^0}
\def\RABb{\bar R_{AB}^0}
\begin{center}
\rule{0cm}{2cm}
{\bf \LARGE Employing internal multiples in time-lapse seismic monitoring, using the Marchenko method}\\
\rule{0cm}{1cm}\\
Kees Wapenaar and Johno van IJsseldijk
\end{center}

\hspace{0cm}\textbf{Summary}\\

Time-lapse seismic monitoring aims at resolving changes in a producing reservoir from changes in the reflection response.
When the changes in the reservoir are very small, the changes in the seismic response can become too small to be reliably detected. In theory, multiple reflections can be used to improve the
detectability of traveltime changes: a wave that propagates several times down and up through a reservoir layer will undergo a larger time shift due to reservoir changes than a primary reflection.
Since we are interested in monitoring very local changes (usually in a thin reservoir layer), it would be advantageous if we could identify the reservoir-related internal multiples
in the complex reflection response of the entire subsurface. We introduce a Marchenko-based method to isolate these multiples from the complete reflection response
and illustrate the potential of this method with numerical examples.

\pagebreak

\section{Introduction}

Time-lapse seismic monitoring aims at resolving changes in a producing reservoir from changes in the reflection response.
Time-lapse changes in the reflection response can consist of (angle-dependent) amplitude changes \citep{Landro2001GEO}, traveltime changes \citep{Landro2004GEO}, or a combination of the two.
When the changes in the reservoir are very small, the changes in the seismic response can become too small to be reliably detected. In theory, multiple reflections can be used to improve the
detectability of traveltime changes: a wave that propagates several times down and up through a reservoir layer will undergo a larger time shift due to reservoir changes than a primary reflection.
This is akin to the underlying principle of coda-wave interferometry \citep{Snieder2002Science, Gret2006GRL}, which employs time-lapse changes in the coda of a multiply-scattered signal to estimate
changes in the background velocity. Since we are interested in monitoring very local changes (usually in a thin reservoir layer), it would be advantageous if we could identify the reservoir-related internal multiples
in the complex reflection response of the entire subsurface. The aim of this paper is to introduce a Marchenko-based method to isolate these multiples from the complete reflection response.

\section{A numerical time-lapse experiment}

\begin{figure}[h]
\centerline{\epsfysize=10. cm \epsfbox{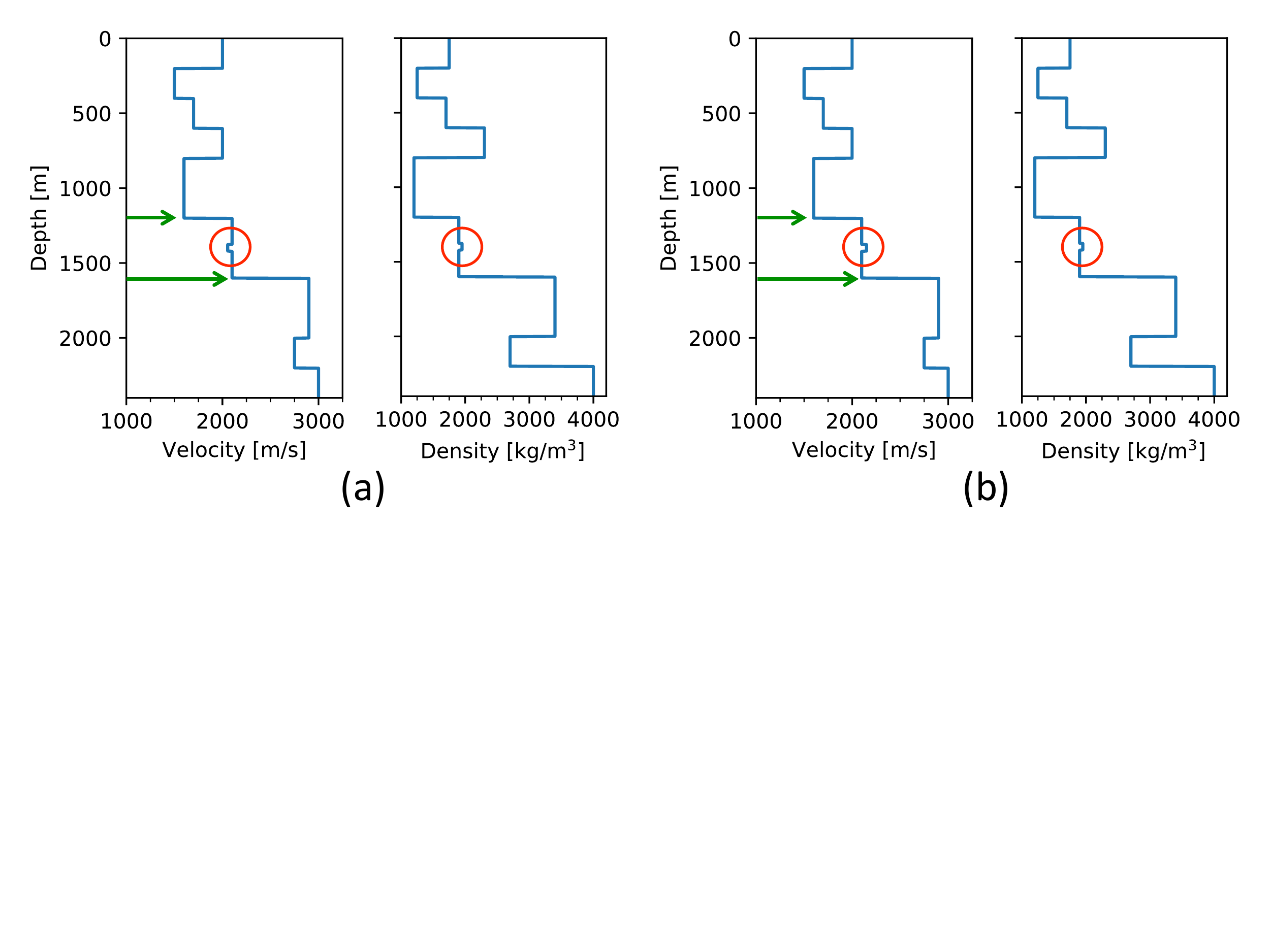}}
\vspace{-5cm}
\caption{\small Velocities and densities of a horizontally layered medium. (a) Baseline state. (b) Monitor state. 
}\label{Fig1}
\end{figure}

\begin{figure}
\vspace{-.3cm}
\centerline{\epsfysize=12. cm \epsfbox{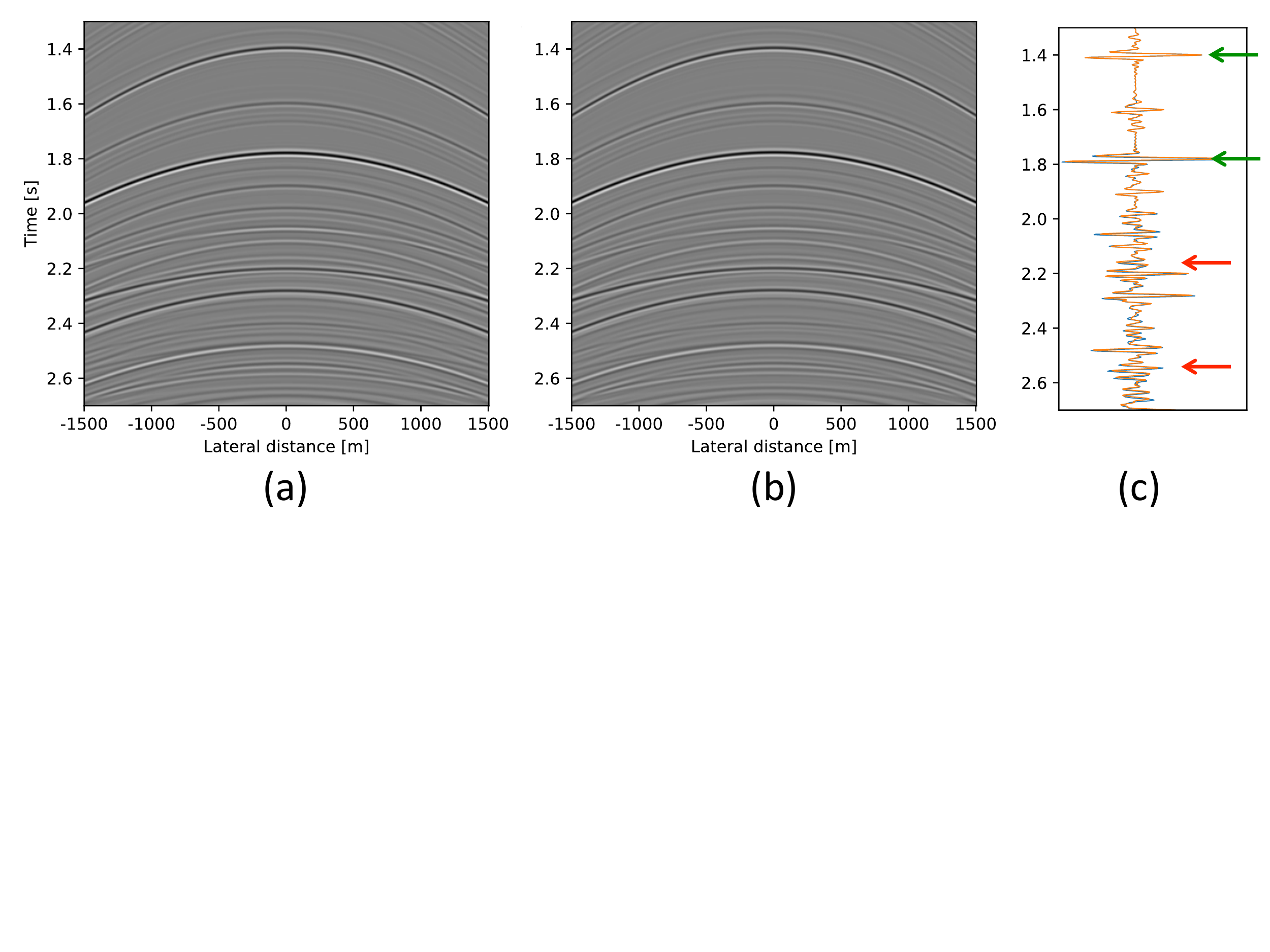}}
\vspace{-6cm}
\caption{\small (a) Baseline reflection response $R(\bxR,\bxS,t)$. (b) Monitor reflection response $\bar R(\bxR,\bxS,t)$. (c) Overlay of central traces of (blue) baseline and (orange) monitor  responses.  
}\label{Fig2}
\end{figure}

Consider a horizontally layered medium, of which the velocities and densities in the baseline and monitor states are shown in Figure \ref{Fig1}. The reservoir layer is encircled.
The thickness of the reservoir layer is 45 m   and the velocities of this layer in the baseline and monitor states are 2055 m/s and 2150 m/s, respectively (the densities are the same in both states). 
Hence, the traveltime shift for primary reflections from interfaces below the reservoir is $-1.94$ ms. The green arrows indicate two strong reflectors above and below the reservoir, at 1200 m and 1600 m, respectively.
Figure \ref{Fig2} shows the baseline and monitor reflection responses $R(\bxR,\bxS,t)$ and $\bar R(\bxR,\bxS,t)$, respectively. Here $\bxS$ and $\bxR$ are the source and receiver coordinates and $t$ denotes time.
The green arrows in Figure \ref{Fig2}(c) indicate the primary reflections of the two reflectors indicated in Figure \ref{Fig1}.
Note that the traveltime shift of the reflector below the reservoir is hardly detectable. Multiples between these reflectors, which have larger traveltime shifts, cannot be identified (they should occur at the traveltimes indicated by the red arrows). 

\section{Isolating the target response, using the Marchenko method}

We define the target zone as the region between two relatively strong reflectors surrounding the reservoir layer.
We propose a two-step method to isolate the target response, including internal multiples between the top and bottom of the target zone, from the complete reflection response.

\begin{figure}[b]
\vspace{-.5cm}
\centerline{\epsfysize=12. cm \epsfbox{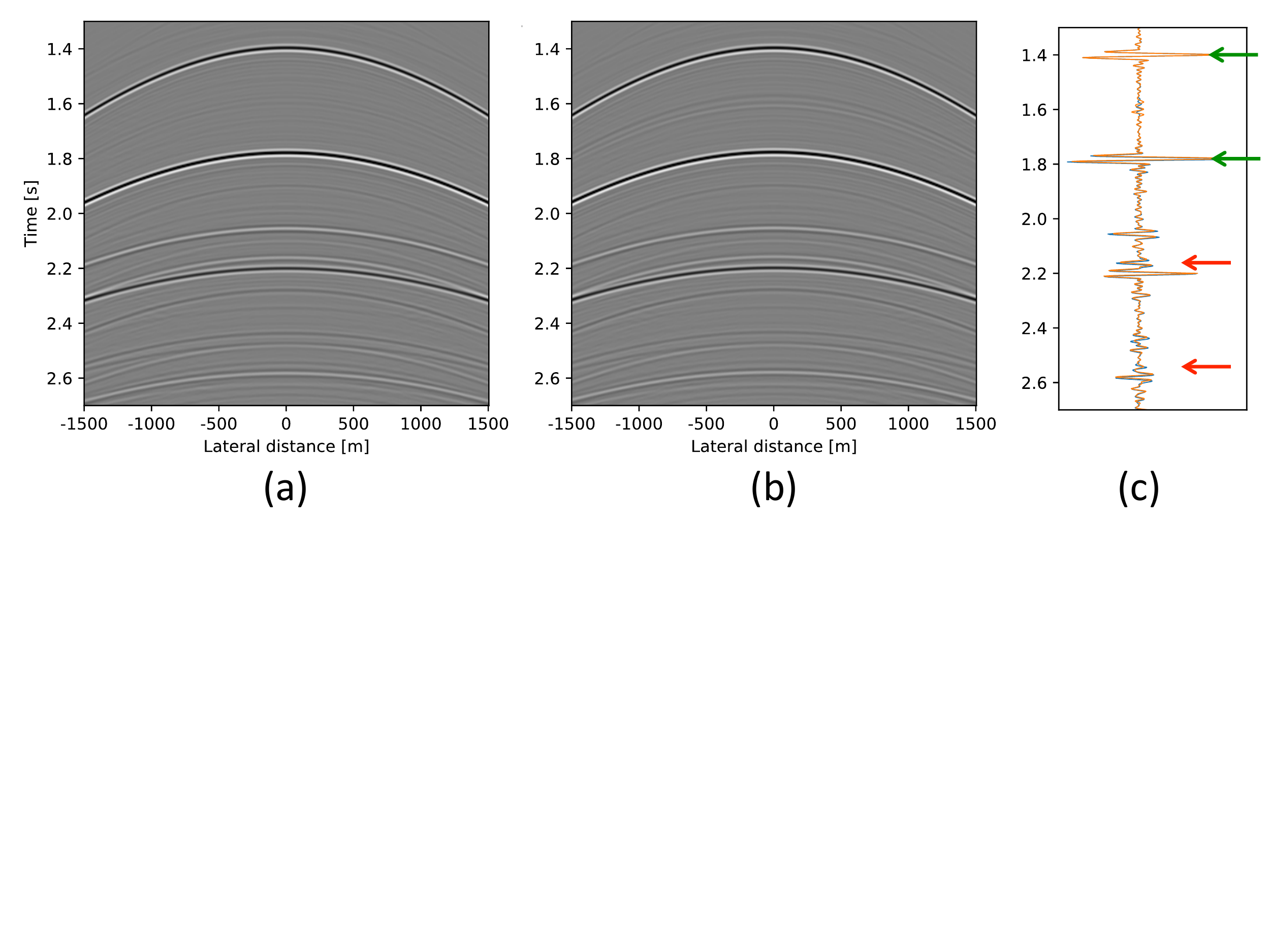}}
\vspace{-6.cm}
\caption{\small Results of removing the overburden response from the baseline and monitor data. (a) $\RAA(\bxR,\bxS,t)$. (b)  $\RAAb(\bxR,\bxS,t)$. (c) Overlay of central traces of (blue) baseline and (orange) monitor  responses.  
}\label{Fig3}
\end{figure}

{\it Step 1: Removing the overburden response.}
We start with the baseline reflection response $R(\bxR,\bxS,t)$ at the acquisition surface $\setdDR$.
We define a focus level $\setdDA$ at a small distance above the target zone.
With the Marchenko method \citep{Wapenaar2014GEO, Slob2014GEO}, using a smooth model of the overburden (the medium between $\setdDR$ and $\setdDA$), 
we retrieve the downgoing and upgoing Green's functions $G^+(\bx',\bxS,t)$ and $G^-(\bx,\bxS,t)$, respectively, 
with observation points $\bx'$ and $\bx$  at $\setdDA$.
These Green's functions are related to each other via 
\begin{equation}\label{eq1}
G^-(\bx,\bxS,t)=\int_{\setdDA} \RA(\bx,\bx',t)*G^+(\bx',\bxS,t)\rmd\bx',
\end{equation}
where $*$ denotes convolution and  $\RA(\bx,\bx',t)$ is the reflection response at $\setdDA$ of the medium below this surface.
Assuming the Green's functions $G^+(\bx',\bxS,t)$ and $G^-(\bx,\bxS,t)$ are retrieved for many source positions $\bxS$ at $\setdDR$, the redatumed reflection response 
$\RA(\bx,\bx',t)$ can be resolved from equation (\ref{eq1}) by multidimensional deconvolution (MDD).
To facilitate a good comparison with the original response $R(\bxR,\bxS,t)$ at the acquisition surface $\setdDR$, we project $\RA(\bx,\bx',t)$ to this surface, according to
\begin{equation}\label{eq2}
\RAA(\bxR,\bxS,t)=\int_{\setdDA}\int_{\setdDA}G_\rmd(\bxR,\bx,t)* \RA(\bx,\bx',t)*G_\rmd(\bx',\bxS,t)\rmd\bx\rmd\bx'
\end{equation}
 \citep{Meles2016GEO, Neut2016GEO}. Here $G_\rmd(\bx',\bxS,t)$ is the (flux-normalised) direct-wave Green's function between $\setdDR$ and $\setdDA$. By defining it 
in the overburden model that is also used for Marchenko redatuming, traveltime errors of the Marchenko redatuming are compensated for by this projection. 
Applying a similar procedure to the monitor reflection response $\bar R(\bxR,\bxS,t)$, using the same overburden model, yields  $\RAAb(\bxR,\bxS,t)$.
Figure \ref{Fig3} shows the result of applying this procedure to the responses in Figure \ref{Fig2}. Note that the responses are cleaner. 
However, the multiples between the top and bottom reflectors of the target zone, indicated by the red arrows in Figure \ref{Fig3}(c), are still contaminated by the primaries from deeper reflectors.
This will be improved further in the next step.

\begin{figure}[t]
\vspace{-.5cm}
\centerline{\epsfysize=12. cm \epsfbox{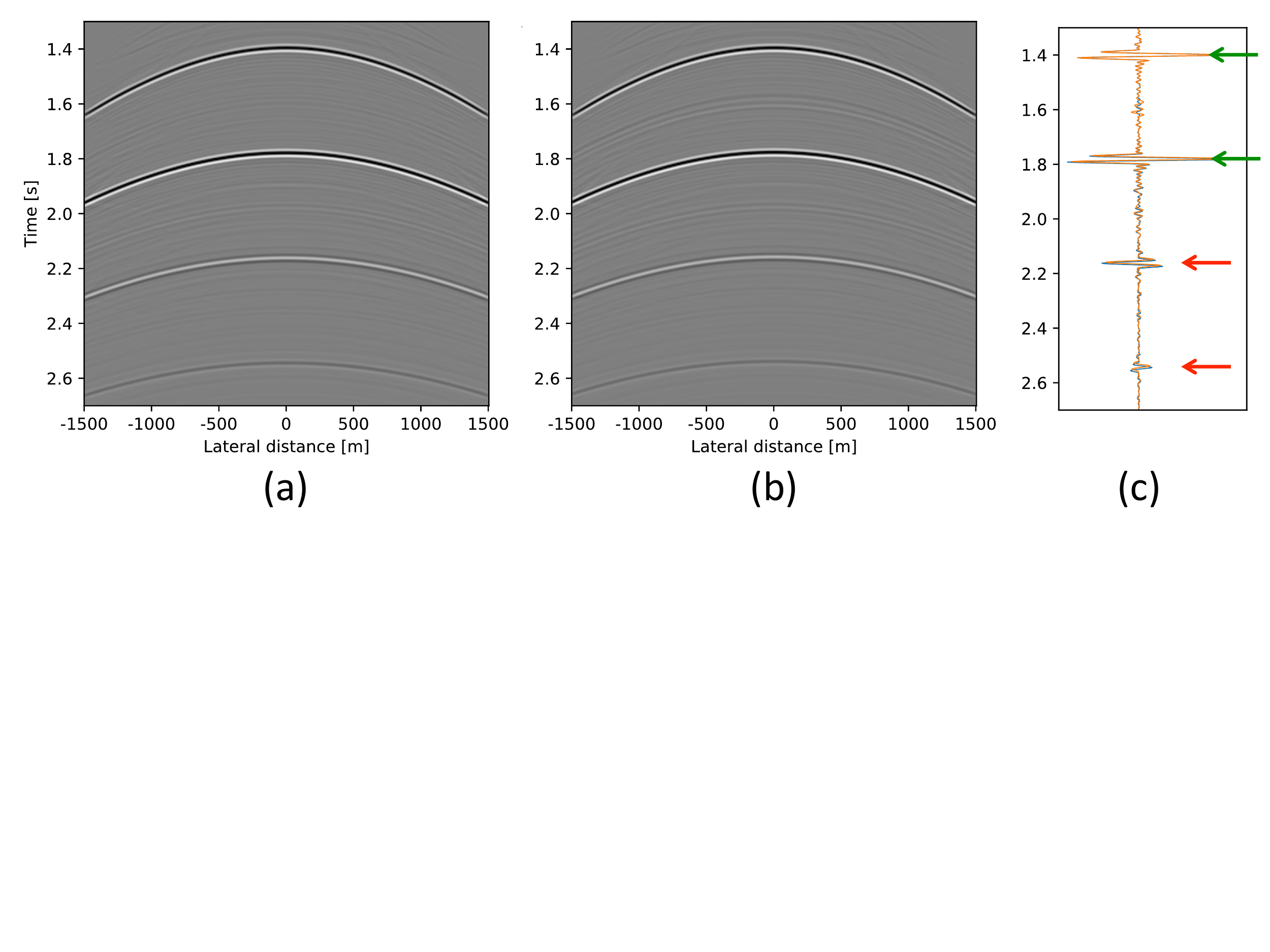}}
\vspace{-6cm}
\caption{\small Results of removing the underburden response from the baseline and monitor data. (a) $\RAB(\bxR,\bxS,t)$. (b)  $\RABb(\bxR,\bxS,t)$. (c) Overlay of central traces of (blue) baseline and (orange) monitor  responses.  
}\label{Fig4}
\end{figure}

{\it Step 2: Removing the underburden response.}
We continue with the baseline response $\RAA(\bxR,\bxS,t)$ of the medium below $\setdDA$, projected to the acquisition surface $\setdDR$.
We define a focus level $\setdDB$ at a small distance below the target zone.
With the Marchenko method, using a smooth model of the overburden and target zone (the medium between $\setdDR$ and $\setdDB$),
we retrieve the downgoing and upgoing focusing functions $f_1^+(\bxS,\bx,t)$ and $f_1^-(\bxR,\bx,t)$, respectively, 
with the focal point $\bx$ at $\setdDB$.
These focusing functions are related to each other via 
\begin{equation}\label{eq3}
f_1^-(\bxR,\bx,t)=\int_{\setdDR} \RAB(\bxR,\bxS,t)*f_1^+(\bxS,\bx,t)\rmd\bxS
\end{equation}
\citep{Wapenaar2018JGR}, where $\RAB(\bxR,\bxS,t)$ is the reflection response at $\setdDR$ of the medium between $\setdDA$ and $\setdDB$ (i.e., the target zone). 
Assuming the focusing functions $f_1^+(\bxS,\bx,t)$ and $f_1^-(\bxR,\bx,t)$  are retrieved for many focal point positions $\bx$ at $\setdDB$, the reflection response 
$\RAB (\bxR,\bxS,t) $ can be resolved from equation (\ref{eq3}) by MDD.
Applying a similar procedure to the monitor reflection response $\RAAb(\bxR,\bxS,t)$, using the same overburden and target model, yields  $\RABb(\bxR,\bxS,t)$.

\begin{figure}
\vspace{-.3cm}
\centerline{\epsfysize=10. cm \epsfbox{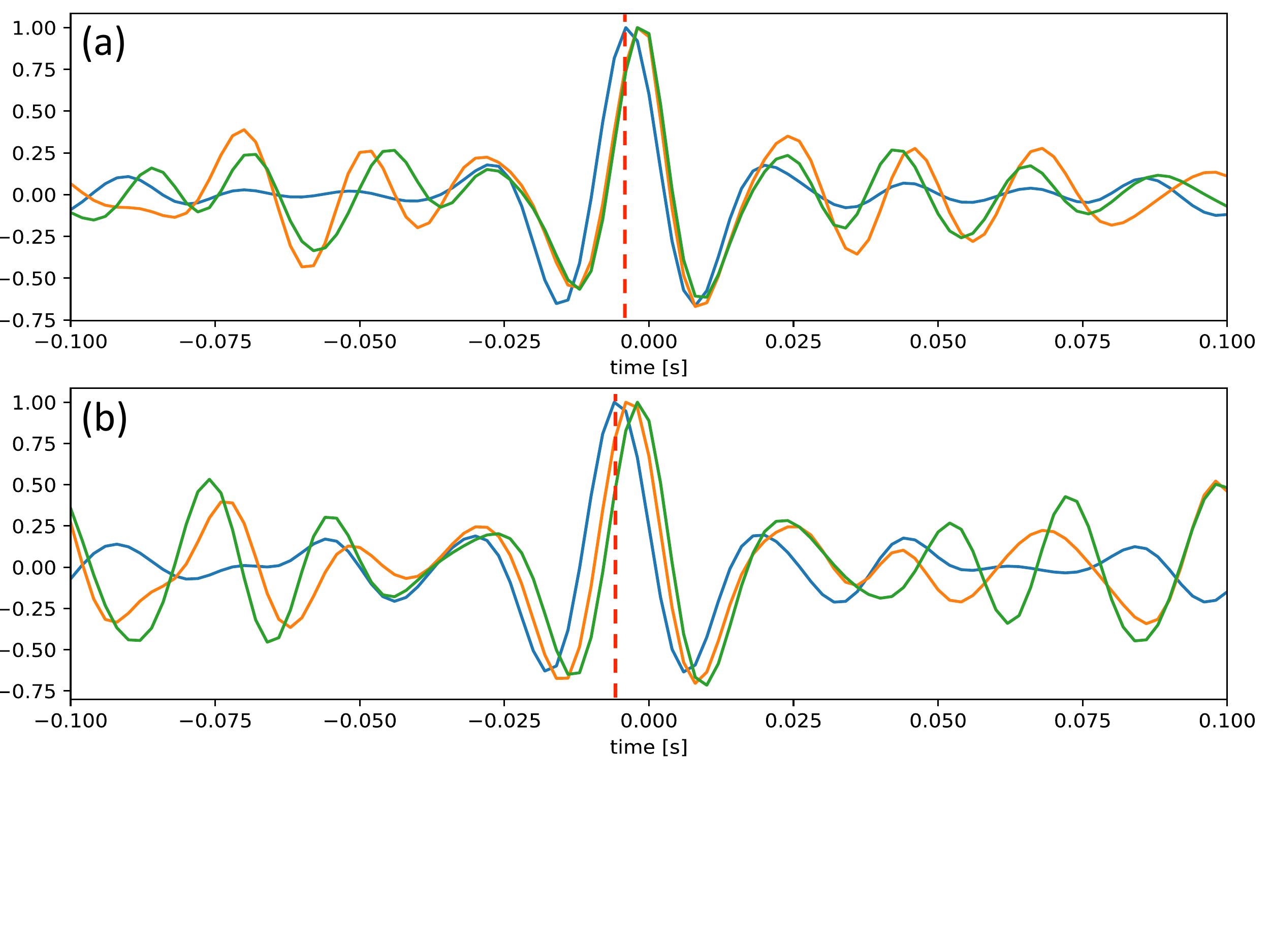}}
\vspace{-2.5cm}
\caption{\small Cross-correlations of the baseline and monitor responses in a time window (a) from 2.0 to 2.2 s around the first multiple and (b) from 2.4 to 2.6 s around the second multiple. 
The traveltime shifts inferred from the blue curves ($-4$ ms and $-6$ ms, respectively) correspond accurately with the expected time shifts for the first and second multiple 
(i.e., two times and three times, respectively, the expected time shift of $-1.94$ ms for the primary).
}\label{Fig5}
\end{figure}

Figure \ref{Fig4} shows the result of applying this procedure to the responses in Figure \ref{Fig3}. 
Note that the responses are again cleaner. In particular, the multiples between the top and bottom reflectors of the target zone, indicated by the red arrows in Figure \ref{Fig4}(c), are now clearly identifiable.
The subtle shifts between the baseline and monitor responses are now also visible.
Coda-wave interferometry can now be used to estimate the changes in the reservoir. To this end we apply cross-correlation of the baseline and monitor responses 
in a time window from 2.0 to 2.2 s, around the first multiple, see Figure \ref{Fig5}(a).
The green curve is the result obtained from the original data (Figure \ref{Fig2}(c)), the orange curve from the data with the overburden response removed (Figure \ref{Fig3}(c)), and the blue 
curve from the data with the overburden and underburden responses removed (Figure \ref{Fig4}(c)). Note that from the blue curve in Figure \ref{Fig5}(a) we infer a time shift of $-4$ ms.
The expected time shift for this multiple is twice the expected time shift of $-1.94$ ms for the primary, hence, the retrieved time shift is quite accurate.
Figure  \ref{Fig5}(b) shows the cross-correlations in a time window from 2.4 to 2.6 s, around the second multiple. The time shift inferred from the blue curve is  $-6$ ms, which corresponds accurately with three times the 
expected time shift of $-1.94$ ms for the primary.

\section{Conclusions}

We have proposed a Marchenko-based procedure to remove the overburden and underburden responses from the seismic reflection response of a producing reservoir.
This method isolates the response of a target zone, including the multiple reflections between the top and bottom reflectors of the target zone.
Since these multiples propagate several times down and up through the reservoir layer (which is included in the target zone), they are more sensitive to time-lapse changes in the reservoir than primaries. Hence,
these multiples can be used to infer time-lapse changes in the reservoir.

To infer small traveltime changes in time-lapse seismic data, a high degree of repeatability  is required. 
When the baseline and monitor surveys are carried out with different acquisition conditions (different sources and/or receivers, different source wavelets, etc.), 
the differences between these surveys due to changes in acquisition may be larger than those due to changes in the reservoir.
We note that the proposed method (in particular the MDD process) has the potential to reduce the acquisition imprint on the time-lapse response. This is subject of current investigations.

\section{Acknowledgements}

This research was funded by the European Research Council (ERC) under the European Union's Horizon 2020 research and innovation programme (grant agreement No: 742703).


\begin{thebibliography}{9}
\providecommand{\natexlab}[1]{#1}

\bibitem[Gr\^et et~al., 2005]{Gret2006GRL}
Gr\^et, A., Snieder, R., Aster, R.C. and Kyle, P.R. [2005] Monitoring rapid
  temporal change in a volcano with coda wave interferometry.
\newblock {\it Geophysical Research Letters}, \textbf{32}, L06304.

\bibitem[Landr{\o}, 2001]{Landro2001GEO}
Landr{\o}, M. [2001] Discrimination between pressure and fluid saturation
  changes from time-lapse seismic data.
\newblock {\it Geophysics}, \textbf{66}(3), 836--844.

\bibitem[Landr{\o} and Stammeijer, 2004]{Landro2004GEO}
Landr{\o}, M. and Stammeijer, J. [2004] Quantitative estimation of compaction
  and velocity changes using 4{D} impedance and traveltime changes.
\newblock {\it Geophysics}, \textbf{69}(4), 949--957.

\bibitem[Meles et~al., 2016]{Meles2016GEO}
Meles, G.A., Wapenaar, K. and Curtis, A. [2016] Reconstructing the primary
  reflections in seismic data by {M}archenko redatuming and convolutional
  interferometry.
\newblock {\it Geophysics}, \textbf{81}(2), Q15--Q26.

\bibitem[Van~der Neut and Wapenaar, 2016]{Neut2016GEO}
Van~der Neut, J. and Wapenaar, K. [2016] Adaptive overburden elimination with
  the multidimensional {M}archenko equation.
\newblock {\it Geophysics}, \textbf{81}(5), T265--T284.

\bibitem[Slob et~al., 2014]{Slob2014GEO}
Slob, E., Wapenaar, K., Broggini, F. and Snieder, R. [2014] Seismic reflector
  imaging using internal multiples with {M}archenko-type equations.
\newblock {\it Geophysics}, \textbf{79}(2), S63--S76.

\bibitem[Snieder et~al., 2002]{Snieder2002Science}
Snieder, R., Gr\^{e}t, A., Douma, H. and Scales, J. [2002] Coda wave
  interferometry for estimating nonlinear behavior in seismic velocity.
\newblock {\it Science}, \textbf{295}, 2253--2255.

\bibitem[Wapenaar and Staring, 2018]{Wapenaar2018JGR}
Wapenaar, K. and Staring, M. [2018] Marchenko-based target replacement,
  accounting for all orders of multiple reflections.
\newblock {\it Journal of Geophysical Research}, \textbf{123}, 4942--4964.

\bibitem[Wapenaar et~al., 2014]{Wapenaar2014GEO}
Wapenaar, K., Thorbecke, J., van~der Neut, J., Broggini, F., Slob, E. and
  Snieder, R. [2014] Marchenko imaging.
\newblock {\it Geophysics}, \textbf{79}(3), WA39--WA57.

\end{thebibliography}
\end{document}